\documentclass[12pt]{article}

\usepackage{graphicx}
\usepackage{amsfonts}
\usepackage[top=2.75cm, bottom=2.75cm, left=2.75cm, right=2.75cm]{geometry} 

\newtheorem{proposition}{Proposition}

\newtheorem{theorem}[proposition]{Theorem}

\def\+{{+\!\!\!+}} 
\def\pp{\mbox{\tiny${}_{\stackrel\+ =}$}}

\def\d{\partial}

\def\P{\Phi}

\def\p{\psi} 
\def\F{\Psi} 
\def\e{\varepsilon}

\def\th{\theta}

\def\pmb#1{\setbox0=\hbox{#1}% 
\kern.0em\copy0\kern-\wd0 
\kern-.04em\copy0\kern-\wd0 
\kern.08em\copy0\kern-\wd0 
\kern-.04em\raise.0433em\box0 }         %poor man's bold macro (TexBook) 
\def\rank{\textstyle{\rm{rank}}} 
 
\newcommand{\nc}{\newcommand} 
\nc{\beq}{\begin{equation}} 
\nc{\eeq}[1]{\label{#1}\end{equation}} 
\nc{\ber}{\begin{eqnarray}} 
\nc{\eer}[1]{\label{#1}\end{eqnarray}} 
\nc{\pek}[1]{\cite{#1}} 
\nc{\enr}[1]{(\ref{#1})} 
\nc{\kal}[1]{{\cal{#1}}} 
\nc{\dott}{\;\cdot\;} 
\nc{\coker}{\mathrm{coker}}
\nc{\ie}{{\it i.e.}}
\nc{\eg}{{\it e.g.}}
\newcommand{\Section}[1]{\section{#1} \setcounter{equation}{0}}

\def\0 {\nonumber}

\begin{document} 
\setcounter{page}{0}
\newcommand{\inv}[1]{{#1}^{-1}} %inverse 
\renewcommand{\theequation}{\thesection.\arabic{equation}} 
\newcommand{\be}{\begin{equation}} 
\newcommand{\ee}{\end{equation}} 
\newcommand{\bea}{\begin{eqnarray}} 
\newcommand{\eea}{\end{eqnarray}} 
\newcommand{\re}[1]{(\ref{#1})} 
\newcommand{\qv}{\quad ,} 
\newcommand{\qp}{\quad .} 

\thispagestyle{empty}
%\begin{titlepage} 
%\title{} 
\begin{flushright} \small
UUITP-21/05 \\ HIP-2005-50/TH \\  YITP-SB-05-40 \\  NSF-KITP-05-106 \\
\end{flushright}
\smallskip
\begin{center} \LARGE
{\bf Generalized K\"ahler manifolds and off-shell supersymmetry}
 \\[12mm] \normalsize
{\bf Ulf~Lindstr\"om$^{a,b}$, Martin Ro\v cek$^{c}$, Rikard von Unge$^{d}$, and Maxim Zabzine$^{a,e}$} \\[8mm]
 {\small\it
$^a$Department of Theoretical Physics 
Uppsala University, \\ Box 803, SE-751 08 Uppsala, Sweden \\
~\\
$^b$HIP-Helsinki Institute of Physics, University of Helsinki,\\
P.O. Box 64 FIN-00014  Suomi-Finland\\
~\\
$^c$C.N.Yang Institute for Theoretical Physics, Stony Brook University, \\
Stony Brook, NY 11794-3840,USA\\
~\\
$^{d}$Institute for Theoretical Physics, Masaryk University, \\ 61137 Brno, Czech Republic 
\\~\\
$^{e}$Kavli Institute of Theoretical Physics, University of California,\\
Santa Barbara, CA 93106 USA
}
\end{center}
\vspace{10mm}
\centerline{\bfseries Abstract} \bigskip

\noindent  
We solve the long standing problem of finding an off-shell supersymmetric 
formulation for a general $N=(2,2)$ nonlinear two dimensional sigma model. 
Geometrically the problem is equivalent to proving
the existence of special coordinates; these correspond to particular superfields
that allow for a superspace description.
We construct and explain the geometric significance of the 
generalized K\"ahler potential for any generalized K\"ahler manifold; this
potential is the superspace Lagrangian.
\eject
\normalsize

%\addtocontents{toc}
%\tableofcontents

%\end{titlepage}

\section{Introduction}

Recently general $N=(2,2)$ supersymmetric sigma models have  
attracted considerable attention; the renewed interest comes both from physics and  
mathematics.
The physics is related to compactifications with NS-NS
fluxes, whereas the mathematics is associated with  
generalized complex geometry,
in particular, generalized K\"ahler geometry, which is precisely 
the geometry of the target space of $N=(2,2)$  
supersymmetric sigma models.

The general $N=(2,2)$ sigma model originally described in 
 \cite {Gates:1984nk} has been studied extensively in the physics literature. 
However, until now an $N=(2,2)$ off-shell  
supersymmetric formulation has not been known in the general case.
At the physicist's level of rigor, a description in terms $N=(2,2)$  
superfields would imply the existence of a single function that encodes the local  
geometry--a generalized K\"ahler potential.
Geometrically the problem of $N=(2,2)$ off-shell supersymmetry  
amounts to the proper understanding of certain natural local coordinates and the 
generalized K\"ahler potential.

In the present work we resolve the issue of what constitutes a  
complete description of the target space geometry of a general
$N=(2,2)$ sigma model. We show that the full set of fields
consists of chiral, twisted chiral and semichiral fields. This was
was a natural guess after semichiral superfields were discovered
in \cite{Buscher:1987uw}, and was explicitly conjectured by
Sevrin and Troost \cite{Sevrin:1996jr};
however, in \cite{Bogaerts:1999jc}, which contains many useful
and interesting results, the erroneous conclusion that this is
not the case was reached.

The bulk of the paper is devoted to the proof that  
certain local coordinates for generalized K\"ahler geometry exist. From the point of view  
of $N=(2,2)$ supersymmetry these coordinates are natural and correspond
to the basic superfield ingredients of the model.

The paper is organized as follows. In Section \ref{GKG} we review the  
general $N=(2,2)$ sigma model and describe the generalized K\"ahler geometry.   
Section \ref{OFF} states the problem of off-shell supersymmetry and explains what should  
be done to solve it. In Section \ref{POISSON} we describe three relevant Poisson  
structures and their symplectic foliations, and identify coordinates  
adapted to these foliations. For the sake of clarity in Section \ref{COKER} we start with a  
special case when $\ker [J_+, J_-] =\emptyset$. In this case we show that the  
correct coordinates exist and we explain the existence of the generalized K\"ahler  
potential. Next, in Section \ref{GENERAL} we extend our results to the general case. 
Finally, in Section \ref{SUMMARY} we summarize our results and explain some open problems.

{\bf Warning to mathematicians:} Due to our background  
as physicists, we like to work in local coordinates with all indices written out. However,  
all expressions can be written in coordinate free form, except when we discuss the  
specific local coordinates in Sections \ref{COKER} and \ref{GENERAL}; however, 
even these local coordinates are merely a convenience, and an appropriate global
reformulation of our results certainly exists.

\section{Generalized K\"ahler geometry}
\label{GKG}

In this section we review the results on general $N=(2,2)$  
supersymmetric sigma models
from the original work \cite{Gates:1984nk} (some of
these results were found independently in 
\cite{Curtright:1984dz,Howe:1985pm}). 
We define our notation and introduce some relevant  
concepts.

We start from the general $N=(1,1)$  sigma model 
written in $N=(1,1)$ superfields
(see Appendix A for our conventions)
\beq
S\propto \int_\Sigma d^2\sigma\,d^2\theta\,\,D_+\Phi^\mu D_- \Phi^\nu (g_{\mu\nu} 
(\Phi)
+ B_{\mu\nu}(\Phi)) ~.
\eeq{actionB}
The action (\ref{actionB}) is manifestly supersymmetric under the usual
supersymmetry transformations
\beq
\delta_1(\epsilon)  \Phi^\mu  = -i (\epsilon^+ Q_+ + \epsilon^- Q_-)  
\Phi^\mu ~,
\eeq{susyI}
which form the standard supersymmetry algebra
\beq
[ \delta_1(\epsilon_1), \delta_1(\epsilon_2)] \Phi^\mu =
-2i \epsilon_1^+ \epsilon_2^+ \d_\+ \Phi^\mu
-2i \epsilon_1^- \epsilon_2^- \d_= \Phi^\mu ~.
\eeq{susyIalgebra}
We may look for additional supersymmetry transformations of the form \cite{Gates:1984nk}
\beq
\delta_2(\epsilon) \Phi^\mu
=\epsilon^+ D_+ \Phi^\nu J^\mu_{+\nu}(\Phi)
+ \epsilon^- D_- \Phi^\nu J^\mu_{-\nu}(\Phi)~ .
\eeq{secsupfl}
Classically the ansatz (\ref{secsupfl}) is unique for dimensional  
reasons.
The action (\ref{actionB}) is invariant under the  
transformations
(\ref{secsupfl}) provided that
\beq
J_{\pm\rho}^\mu g_{\mu\nu} = - g_{\mu\rho} J^\mu_{\pm\nu}
\eeq{bla}
and
\beq
\nabla^{(\pm)}_\rho J^\mu_{\pm\nu} \equiv J^\mu_{\pm\nu,\rho} +
\Gamma^{\pm\mu}_{\,\,\rho\sigma} J^\sigma_{\pm\nu} - \Gamma^{\pm 
\sigma}_{\,\,\rho\nu}
J^\mu_{\pm\sigma}=0~,
\eeq{nablJH}
where the two affine connections
\beq
\Gamma^{\pm\mu}_{\,\,\rho\nu} = \Gamma^{\mu}_{\,\,\rho\nu} \pm g^ 
{\mu\sigma} H_{\sigma\rho\nu}
\eeq{defaffcon}
have torsion determined by the field strength of $B_{\mu\nu}(\Phi)$:
\beq
H_{\mu\rho\sigma} = \frac{1}{2}
(B_{\mu\rho,\sigma} + B_{\rho\sigma,\mu} + B_{\sigma\mu,\rho})~.
\eeq{Hdefin}
Indeed the functional (\ref{actionB}) can be rewritten in terms of
an extension of $H$ to a ball whose boundary is the surface $\Sigma$
modulo the usual arguments that apply to the bosonic WZW-term, namely 
$[H] \in H^3(M, \mathbb{Z})$.

Next  we impose the standard on-shell $N=(2,2)$  
supersymmetry algebra: The first
supersymmetry transformations (\ref{susyI}) and the second  
supersymmetry transformations (\ref{secsupfl})
automatically commute
\beq
[\delta_2(\epsilon_1), \delta_1(\epsilon_2)]\Phi^\mu = 0~.
\eeq{commtwosusy}
The commutator of two second supersymmetry transformations,
\ber
\nonumber[\delta_2(\epsilon_1), \delta_2(\epsilon_2)]\Phi^\mu \!\!&=&\! \!
2i \epsilon_1^+ \epsilon_2^+ \d_\+ \Phi^\lambda
( J^\mu_{+\nu}J^\nu_{+\lambda}) +2i \epsilon_1^- \epsilon_2^- \d_=   
\Phi^\lambda
( J^\mu_{-\nu}J^\nu_{-\lambda}) \\ 
\nonumber&&\!\! -\, \epsilon_1^+ \epsilon_2^+ D_+\Phi^\lambda D_+\Phi^\rho
{\cal N}^\mu_{\,\,\lambda\rho}(J_+) -  \epsilon_1^- \epsilon_2^- D_- 
\Phi^\lambda D_-\Phi^\rho
{\cal N}^\mu_{\,\,\lambda\rho}(J_-) \\ 
&& \!\! +\, (\epsilon_1^+ \epsilon_2^- + \epsilon^-_1 \epsilon_2^+) (J_{+\nu} 
^\mu J_{-\lambda}^\nu -
J_{-\nu}^\mu J_{+\lambda}^\nu) (D_+ D_- \Phi^\lambda + \Gamma^{- 
\lambda}_{\,\,\sigma
\nu} \,D_+\Phi^\sigma D_-\Phi^\nu )~,\nonumber\\
\eer{secondsusy}
should satisfy the same algebra as the first (\ref{susyIalgebra}), \ie,
\beq
[\delta_2(\epsilon_1), \delta_2(\epsilon_2)]\Phi^\mu =-2i \epsilon_1^+  
\epsilon_2^+ \d_\+ \Phi^\mu
- 2i \epsilon_1^- \epsilon_2^- \d_=  \Phi^\mu~.
\eeq{secondlikefirst}
In (\ref{secondsusy}),
${\cal N} (J_{\pm})$ is the Nijenhuis tensor defined by
\beq
{\cal N}^{\rho}_{\,\,\mu\nu} (J) = J^\rho_\lambda\d_{[\nu}J^\lambda_{\mu]}+
 \d_\lambda J^\rho_{[\nu}J^\lambda_{\mu]}~.
\eeq{intergpl11}
The field equations that follow from the action (\ref{actionB}) are
\beq
D_+ D_- \Phi^\mu + \Gamma^{-\mu}_{\,\,\rho\sigma} \,D_+\Phi^\rho D_- 
\Phi^\sigma = 0~.
\eeq{eqsmot}
The first two lines of (\ref{secondsusy}) are purely kinematical, \ie, are independent
of the form of the action; the last
line involves the field equations (\ref{eqsmot}), and follows after imposing (\ref{nablJH}).

The algebra (\ref{secondsusy}) is the usual supersymmetry algebra (\ref{susyIalgebra}) when $J_{\pm}$ obey:
\beq
~J^\mu_{\pm\nu} J^\rho_{\pm\mu} = -\delta ^\rho{}_\nu~,
\eeq{complstruct}
\beq
{\cal N}^{\rho}{}_{\mu\nu} (J_{\pm}) = 0~;
\eeq{intergpl}
the last term in (\ref {secondsusy}) must also vanish; this is automatic on-shell, \ie, when the field equations (\ref{eqsmot}) are satisfied.
Thus the on-shell supersymmetry algebra requires
that $J_{\pm}$ are integrable complex structures that preserve the metric; we may
introduce the forms $\omega_\pm = g J_\pm$, which are {\em not} closed, but satisfy
\beq
H_{\mu\nu\rho} = \mp J^\lambda_{\pm\mu} J^\sigma_{\pm\nu} J^\gamma_ 
{\pm\rho}
(d\omega_\pm )_{\lambda\sigma\gamma}~,
\eeq{veryusefulrelapdkjas}
as follows from (\ref{bla}), (\ref{nablJH}), (\ref{complstruct})  
and (\ref{intergpl}).

This is the full description of the most general $N=(2,2)$ sigma  
model \cite{Gates:1984nk}: The target
manifold $(M, g, J_\pm, H)$
is a bihermitian complex manifold (\ie, there are two  
complex structures
and a metric that is Hermitian with respect to both) and the two complex  
structures must
be covariantly constant with respect to connections that differ by the sign of
the torsion; this torsion is expressed in terms of a closed 3-form that obeys (\ref{veryusefulrelapdkjas}).

This bihermitian geometry was first described in \cite{Gates:1984nk}.  
Subsequently, a different geometric interpretation was given in \cite{Lyakhovich:2002kc},
and more recently, following ideas of Hitchin
 \cite{hitchinCY}, Gualtieri \cite {gualtieri} gave an entirely  
new description of this geometry in terms of generalized
complex structures. This geometry is now  known as generalized K\"ahler
geometry.

\section{$N=(2,2)$ off-shell supersymmetry}
\label{OFF}

In the previous section, the field equations \enr{eqsmot} are needed to close 
the supersymmetry algebra. To write the model in a manifestly
$N=(2,2)$ covariant form, the algebra must close off-shell. As can be seen from
(\ref{secondsusy}), the algebra does close off-shell when the two complex
structures commute \cite{Gates:1984nk}: $[J_+,J_{-}] = 0$.
In this case, both complex structures and the product structure
$\Pi = J_{+}J_{-}$ are integrable and simultaneously diagonalizable.
The manifest $N=(2,2)$ formulation is given in terms of chiral  ($\phi 
$) and twisted chiral ($\chi$)
scalar superfields:
\ber
&&\bar \mathbb{D}_{\pm}\phi=\mathbb{D} _{\pm}\bar\phi=0\cr
&&\bar\mathbb{D}_{+}\chi= \mathbb{D}_{-}\chi=
\mathbb{D}_{+}\bar\chi=\bar\mathbb{D}_{-}\bar \chi=0~,
\eer{bla192920}
where $\mathbb{D}$ is the  $N=(2,2)$ covariant derivative.
The $N=(2,2)$ Lagrangian is a general real function $K(\phi,\bar\phi, 
\chi,\bar\chi)$, defined modulo
(the equivalent of)  a K\"ahler gauge transformation:
$f(\phi,\chi)+\bar f(\bar\phi,\bar\chi)+g(\phi,\bar\chi)+\bar g(\bar 
\phi,\chi)$. This $K$ serves as a
potential both for the metric and for the antisymmetric $B$-field.

When $[J_+,J_{-}] \ne 0$, additional (auxiliary) spinorial $N= 
(1,1)$ fields are needed to close the algebra
 \cite{Lindstrom:2004eh}, \cite{Lindstrom:2004iw}.  The semichiral
$N=(2,2)$ scalar superfields introduced in \cite{Buscher:1987uw}
give rise to such auxiliary fields when they are reduced to $N=(1,1)$ 
superspace. 
A complex left semichiral superfield $\mathbb{X}_L$ obeys
\beq
\bar \mathbb{D}_{+}\mathbb{X}_L= \mathbb{D}_+ \bar\mathbb{X}_L = 0~,
\eeq{left}
and a right semichiral superfield $\mathbb{X}_R$ obeys
\beq
\bar\mathbb{D}_- \mathbb{X}_R = \mathbb{D}_{-}\bar\mathbb{X}_R = 0~.
\eeq{right}
For these multiplets, the $N=(2,2)$ nonlinear sigma model Lagrangian\footnote{In 
 \cite{Buscher:1987uw}, for simplicity, no chiral or twisted chiral multiplets
are considered, and hence $[J_+,J_{-}]$ is invertible.} is the real
function $K(\mathbb{X}_L,\bar\mathbb{X}_L,\mathbb{X}_R,\bar\mathbb{X}_R)$,  
defined modulo $f(\mathbb{X}_L)+\bar f(\bar\mathbb{X}_L)+ g(\mathbb{X}_R)+\bar
g(\bar\mathbb{X}_R)$.
Again, the function $K$ is a potential for the metric and the  
antisymmetric $B$-field \cite{Buscher:1987uw}.
The target space has generalized K\"ahler geometry with  $[J_+,J_{-}]  
\ne 0$ \cite{Lindstrom:2004hi}.
However, before our work, it was not known if all generalized K\"ahler  
geometries with $[J_+,J_{-}] \ne 0$ admit a description in terms of
semichiral superfields.

In \cite{Ivanov:1994ec}, it is shown that the kernel of $[J_+,J_{-}]$  
is parametrized completely by chiral and twisted chiral fields.
This does not answer the question of whether semichiral multiplets  
similarily give a complete description of
the cokernel. The issue has been addressed, \eg, in \cite{Sevrin:1996jq}, 
 \cite{Sevrin:1996jr} and \cite{Grisaru:1997pg}.

The general sigma model Lagrangian containing chiral, twisted  
chiral, and semichiral fields is a
real function
\be
K(\phi,\bar\phi,\chi,\bar\chi,\mathbb{X}_L,\bar\mathbb 
{X}_L, \mathbb{X}_R,\bar\mathbb{X}_R)
\ee 
defined modulo $f(\phi, \chi,\mathbb{X}_L)+g(\phi,
\bar\chi,\mathbb{X}_R )+
\bar f(\bar\phi,\bar\chi,\bar\mathbb{X}_L)+\bar g(\bar\phi,\chi,\bar 
\mathbb{X}_R)$. When there are several multiplets of each kind\footnote{To 
be able to integrate out the auxiliary $N=(1,1)$
spinor superfields, we require an equal number of left and right semichiral superfields 
$\mathbb{X}_L$ and $\mathbb{X}_R$.}, the fields carry indices
\ber
\phi^\alpha, \bar\phi^{\bar\alpha}~,~~\alpha=1\dots d_c&~,~~&
\chi^{\alpha'}, \bar\chi^{\bar \alpha'}~,~~\alpha ' =1\dots d_t~,~~\nonumber\\
\mathbb{X}_L^{a}, \bar\mathbb{X}_L^{\bar a}~,~~a =1\dots d_s&~,~~&
\mathbb{X}_R^{a'},\bar\mathbb{X}_R^{\bar a'}~,~~a'=1\dots d_s~.
\eer{indices}
We will also use the collective notation $\kal{A}:=\{\alpha, \bar 
\alpha\}$,
$\kal{A}':=\{\alpha ', \bar\alpha '\}$, $A:=\{a, \bar a\}$ and $A':=\{a', \bar a'\}$.
To reduce the $N=(2,2)$ action to its $N=(1,1)$ form, we introduce  
the $N=(1,1)$ covariant derivatives
$D$ and extra supercharges $Q$:
\ber
&&D_{\pm}=\mathbb{D}_{\pm}+\bar \mathbb{D}_{\pm}\cr
&&Q_{\pm}=i(\mathbb{D}_{\pm}-\bar \mathbb{D}_{\pm})~.
\eer{twone}
In terms of these, the (anti)chiral, twisted  (anti)chiral  and  
semi (anti)chiral superfields satisfy
\ber
Q_{\pm}\phi = J_c D_{\pm} \phi &~,~~&
Q_\pm \chi = \pm J_t D_\pm \chi~,\cr
Q_+ \mathbb{X}_L= J_s D_+ \mathbb{X}_L&~,~~&
Q_- \mathbb{X}_R = J_s D_- \mathbb{X}_R~,
\eer{QDrels}
where the collective notation is used in the matrices, and where $J_ 
{c}, J_{t}$, and $J_{s}$ are  $2d_{c}, 2d_{t}$, and $2d_{s}$ dimensional 
canonical complex structures of the form
\beq
J = \left( \begin{array}{ll}
i & \,\,\,0\\
0 & -i
\end{array}\right)~.
\eeq{defincanoncsysm}
For  the pair $(\phi,\chi)$
we use the same letters  to denote the $N=(1,1)$
superfields, \ie, the lowest components of the $N=(2,2)$ superfields  
$(\phi,\chi)$.
Each of the semi (anti)chiral fields gives rise to two $N=(1,1)$  
fields:
\ber
&&X_{L}\equiv \mathbb{X}_L|\qquad \F_{L-}\equiv Q_{-}\mathbb{X}_L |\cr
&&X_{R}\equiv \mathbb{X}_R|\qquad \F _{R+}\equiv Q_{+}\mathbb{X}_R |~,
\eer{twone2}
where a vertical bar means that we take the
$\theta^2\propto\theta-\bar\theta$ independent component.
The conditions \enr{QDrels} then also imply
\ber
&&Q_{+}\F_{L-}=J_{s}D_{+}\F_{L-}~,\qquad Q_{-}\F_{L-}= -i\partial_ 
{=}X_{L}\cr
&&Q_{-}\F_{R+}=  J_{s}D_{-}\F_{R+}~,\qquad Q_{+}\F_{R+}=
-i\partial_{\+}X_{R}~.
\eer{QF}

Using the relations \enr{twone}-\enr{QF} we reduce the  $N=(2,2)$
action to its $N=(1,1)$ form according to:
\beq
\int d^2\xi d^2\th d^2\bar\th~ K (\phi^{\kal{A}}, \chi^{\kal{A}'}, 
\mathbb{X}_L^{A},\mathbb{X}_R^{A'})|
=\int d^2\xi \mathbb{D} ^2\bar \mathbb{D}^2 K|=-\frac i 4 \int d^2 
\xi D^{2}Q_{+}Q_{-}
K|~.
\eeq{Act2}
Provided that the matrix
\ber
&K_{LR} \equiv\left(\begin{array}{cc}
K_{ab'} & K_{a\bar{b}'}\cr
K_{\bar{a}b'}& K_{\bar{a} \bar{b}'}\end{array}\right)~.
\eer{HM}
is invertible, the auxiliary spinors $\F_{L-},\F_{R+}$ may be integrated
out leaving us with a
$N=(1,1)$ second order
sigma model action of the type originally discussed in \cite{Gates:1984nk}.
In (\ref{HM}) we use the following notation $K_{ab} \equiv \d_a \d_b K 
$ etc.
From this the metric and antisymmetric
$B$-field may be read off in terms of derivatives of $K$, and from the
form of the second supersymmetry the complex structures $J_{\pm}$  
are determined.
In a basis where the coordinates are arranged in a column as
\ber
\left(\begin{array}{c}
X_{L}^{A}\cr
X_{R}^{A'}\cr
\phi^{\kal{A}}\cr
\chi^{\kal{A}'}\end{array}\right)~,
\eer{Column}
and introducing the notation (suppressing the hopefully obvious index
structure)
\ber
K_{LR}^{-1} &=& (K_{RL})^{-1}~,\cr
C &=& JK-KJ
=\left(\begin{array}{cc}
0 & 2i K\cr
-2i K & 0
\end{array}\right),\cr\cr\cr
A &=& JK+KJ
=\left(\begin{array}{cc}
2i K & 0\cr
0 & -2i K
\end{array}\right),
\eer{commnotation}
the complex structures read \cite{Bogaerts:1999jc}
\ber
J_{+}=
\left(\begin{array}{cccc}
J_s &0&0&0\cr
K_{RL}^{-1}C_{LL} &  K_{RL}^{-1}J_s K_{LR} &
K_{RL}^{-1}C_{Lc}
& K_{RL}^{-1}C_{Lt}\cr
0&0&J_c&0\cr
0&0&0&J_t\end{array}\right)~
\eer{Jplus}
and
\ber
J_{-}=\left(\begin{array}{cccc}
K_{LR}^{-1}J_s K_{RL} & K_{LR}^{-1}C_{RR} &
K_{LR}^{-1}C_{Rc}& K_{LR}^{-1}A_{Rt}\cr
0& J_s&0&0\cr
0&0&J_c&0\cr
0&0&0& -J_t\end{array}\right)~
\eer{Jminus}
where, \eg, $K_{Rc}$ is the matrix of second derivatives along $R 
$- and $c$-directions, etc.
In  Sections \ref{COKER} and \ref{GENERAL}, where we rederive
these expressions from geometrical considerations, we explain  
the notation in greater detail.

Finally, we compute the $N=(1,1)$ Lagrangian; the sum $E=\frac12(g+B)$ of the metric $g$ and 
$B$-field takes on the explicit form:
\ber
E_{LL} &=& C_{LL}K_{LR}^{-1}J_sK_{RL} \cr
E_{LR} &=& J_sK_{LR}J_s + C_{LL}K_{LR}^{-1}C_{RR} \cr
E_{Lc} &=& K_{Lc} + J_s K_{Lc} J_c + C_{LL}K_{LR}^{-1}C_{Rc}\cr
E_{Lt} &=& -K_{Lt} - J_s K_{Lt} J_t + C_{LL}K_{LR}^{-1}A_{Rt}\cr
E_{RL} &=& -K_{RL}J_s K_{LR}^{-1} J_s K_{RL}\cr
E_{RR} &=& -K_{RL}J_s K_{LR}^{-1} C_{RR}\cr
E_{Rc} &=& K_{Rc} - K_{RL}J_s K_{LR}^{-1} C_{Rc}\cr
E_{Rt} &=& -K_{Rt} - K_{RL}J_s K_{LR}^{-1} A_{Rt}\cr
E_{cL} &=& C_{cL}K_{LR}^{-1}J_s K_{RL}\cr
E_{cR} &=& J_c K_{cR} J_s + C_{cL}K_{LR}^{-1}C_{RR}\cr
E_{cc} &=& K_{cc}+J_c K_{cc} J_c + C_{cL}K_{LR}^{-1}C_{Rc}\cr
E_{ct} &=& -K_{ct}-J_c K_{ct}J_t + C_{cL}K_{LR}^{-1}A_{Rt}\cr
E_{tL} &=& C_{tL}K_{LR}^{-1}J_s K_{RL}\cr
E_{tR} &=& J_t K_{tR} J_s + C_{tL}K_{LR}^{-1}C_{RR}\cr
E_{tc} &=& K_{tc} + J_t K_{tc} J_c + C_{tL}K_{LR}^{1}C_{Rc}\cr
E_{tt} &=& -K_{tt} - J_t K_{tt} J_t + C_{tL} K_{LR}^{-1} A_{Rt}
\eer{E}
It is interesting that there are no corrections from chiral and twisted
chiral fields in the semichiral sector (where the results agree with
 \cite{Buscher:1987uw} and \cite{Lindstrom:2004hi}),
whereas in the chiral and twisted chiral sector the semichiral fields  
contribute substantially.

Thus locally all objects ($J_\pm$, $g$, $B$) are given in terms of  
second derivatives of a single real function $K$. By construction,
the present geometry is generalized K\"ahler geometry
and therefore satisfies all the relations from the previous section.
In the rest of the paper we show that (locally) any  
generalized K\"ahler manifold has such a description.

\section{Poisson structures}
\label{POISSON}

In this section we describe three Poisson structures that arise in 
generalized K\"ahler geometry.  We study these Poisson structures as
we will use local coordinates adapted to their foliations. Since the Poisson  
geometry is rather a novel subject to some physicists, we collect some  
basic facts in Appendix C.

We start with the two real Poisson structures
\beq
\pi_\pm \equiv (J_+ \pm J_-) g^{-1}=-g^{-1}(J_+ \pm J_-)^t~,
\eeq{definrealPois}
which were introduced in \cite{Lyakhovich:2002kc} and  later  
rederived by Gualtieri \cite{gualtieri}. We can choose local  
coordinates in a neighborhood of a regular point $x_0$ of 
$\pi_-$ such that\footnote{A regular point $x_0$
of a Poisson structure $\pi$ is a point where the rank of $\pi$ does not vary in
a neighborhod of $x_0$; see Appendix C.} 
\beq
\pi_-^{{\cal A}\mu}= 0~,
\eeq{localcoorPP1}
where ${\cal A}$ label the coordinates along the kernel of $\pi_-$; using 
(\ref{definrealPois}), in these coordinates the complex structures obey
\beq
J^{\cal A}_{+\nu} = J^{\cal A}_{-\nu}~.
\eeq{defsjkeoe9}
Repeating the same argument for $\pi_+$ we get
\beq
J^{\cal A'}_{+\nu} = - J^{\cal A'}_{-\nu}~,
\eeq{blablaN234}
where ${\cal A'}$ label the coordinates along the kernel of $\pi_+$.  Moreover, 
as the combinations $(\pi_+ \pm \pi_-)\propto J_\pm$ are nondegenerate, the Poisson brackets  
defined by $\pi_+$ and $\pi_-$ cannot have common 
Casimir functions\footnote{Casimir functions give the coordinates along which a
Poisson structure is degenerate; see Appendix~C.}
which parametrize the kernels of $\pi_\pm$.
This means that the directions ${\cal A}$ and ${\cal A'}$ do  
not intersect and we can choose coordinates where both the relations 
(\ref{defsjkeoe9}) and (\ref{blablaN234}) hold \cite{Lyakhovich:2002kc}. We 
denote the remaining directions by $A$ and $A'$ 
(for the moment, we do not distinguish $A$ and $A'$).  
Thus we have shown that there exist coordinates, labeled by
$\mu=(A, A', {\cal A}, {\cal A'})$, where
\beq
J_+ = \left( \begin{array}{cccc}
* & * &  * &  *\\
**&*& *& *\\
0 & 0 & J_c & 0\\
0 & 0 & 0& J_t
\end{array} \right)~,\qquad\qquad
J_- = \left( \begin{array}{cccc}
* & * & * & *\\
**&*& *& *\\
0 & 0 & J_c & 0\\
0 & 0 & 0& -J_t
\end{array} \right)~,
\eeq{speccorosl278}
and where $J_c, J_t$ are canonical complex structures defined  
as in
(\ref{defincanoncsysm}).
The existence of these coordinates was originally shown in \cite{Ivanov:1994ec}.
Using Poisson geometry this result is rederived in \cite{Lyakhovich:2002kc}.
We can thus choose  local coordinates adapted to the following
decomposition
\beq
\ker (J_+ - J_-) \oplus \ker (J_+ + J_-) \oplus  \coker [J_+, J_-]~,
\eeq{decompjJJJ}
where we use the property
\beq
[J_+, J_-] = (J_+ - J_-)(J_+ + J_-) = - (J_+ + J_-)(J_+ - J_-)~.
\eeq{decomcomm123}

Another important Poisson structure
\beq
\sigma =  [J_+, J_-] g^{-1}
\eeq{definHitchpoisos}
was introduced in \cite{hitchinP}.  It is related to the real Poisson structures
(\ref{definrealPois}):
\beq
\sigma = \pm (J_+ \mp J_-) \pi_\pm=\mp(J_+\pm J_-)\pi_\mp~.
\eeq{rektooldPoo1}
The identity (\ref{decomcomm123}) implies a relation between the kernels of the three structures
\beq
\ker \sigma = \ker \pi_+ \oplus \ker \pi_-~.
\eeq{kernelsporisoap}
The symplectic leaf for $\sigma$ is $\coker [J_+, J_-]$. The Poisson structure 
$\sigma$ satisfies $J_\pm \sigma J_\pm^t=-\sigma$; this implies that 
in complex coordinates with respect to either $J_\pm$, 
\beq
\sigma = \sigma^{(2,0)} + \bar{\sigma}^{(0,2)}~,
\eeq{sigamdecomshdol}
which implies that the real dimension of the  
symplectic leaves for $\sigma$ is a multiple of $4$ (this was first
proven in \cite{Sevrin:1996jr}).
Indeed, $\sigma$ can be interpreted as the $(2,0)+(0,2)$ 
projection of \eg, $\pi_+$, with respect to either $J=J_\pm$: 
\beq
(1\pm iJ) \sigma (1\pm iJ)^t=\mp2i(1\pm iJ) \pi_+ (1 \pm iJ)^t~.
\eeq{sigmapi}
It turns out that $\sigma^{(2,0)}$ is actually a holomorphic Poisson 
structure \cite{hitchinP}:
\beq
\bar\d \sigma^{(2,0)}=0~,
\eeq{defindholomr}

As discussed above (\ref{decompjJJJ}), we have established 
that along the kernel of $\sigma$, complex coordinates
can be simultaneously chosen for both $J_+$ and $J_-$.  Using
the properties of the cokernel of $\sigma$, in particular 
(\ref{sigamdecomshdol},\ref{defindholomr}), in the next two sections
we find natural coordinates along the symplectic leaf of $\sigma$ as well.  

\section{Structure of $\coker [J_+, J_-]$}
\label{COKER}

To simplify the argument, we first consider
the special case when
$\ker [J_+, J_-] =\emptyset$ on $M$ and $\sigma$ is thus invertible;
this implies $d_c=d_t=0$, and the complex dimension of $M$
is $2d_s$.

Since $\sigma$ is a Poisson structure, the two-form\footnote{This  
two-form was introduced in \cite{Bogaerts:1999jc}; however the authors erroneously
concluded that there exist obstructions to the existence of the 
coordinates that make $\Omega$ constant.}
\beq
\Omega = \sigma ^{-1}~,
\eeq{deftwofrom}
is closed $d\Omega=0$; it also satisfies $J^t_\pm  \Omega J_\pm = - \Omega$.
Choosing complex coordinates with respect to $J_+$,
\beq
J_+ = \left(\begin{array}{cccc}
i &0 & 0 & 0\\
0 &-i & 0 & 0 \\
0 & 0& i & 0 \\
0 & 0 & 0& -i
\end{array}\right) \equiv \left( \begin{array}{cc}
J_s & 0\\
0 & J_s
\end{array} \right),
\eeq{form23-a}
we decompose the symplectic form 
$\Omega$ into its $(2,0)$ and $(0,2)$ parts \cite{Bogaerts:1999jc}
\beq
\Omega = \Omega_+^{(2,0)} + \bar{\Omega}_+^{(0,2)}~.
\eeq{decompomega}
Then $d\Omega=0$ implies
\beq
\d \Omega_+^{(2,0)}=0~,\qquad\bar{\d} 
\Omega_+^{(2,0)}=0~,
\eeq{cloholomomega}
and its complex conjugate expressions with $\d$ ($\bar{\d}$)  
being a holomorphic (antiholomorphic) differential.
Thus $\Omega_+^{(2,0)}$ is a holomorphic symplectic structure and  
according to Darboux's theorem we can choose coordinates 
$\{ q^a, \bar q^{\bar{a}}, p^{a}, \bar p^{\bar a}\},~a=1\dots d_s$ such that
\beq
\Omega_+^{(2,0)} = d q^a \wedge d p^{a}~,\qquad
\bar{\Omega}_+^{(0,2)} = d\bar q^{\bar{a}} \wedge d \bar p^{\bar a }~ .
\eeq{drholomcoord}
These coordinates are compatible with (\ref{form23-a}); the choice of which
coordinates we call $q$ and which we call $p$ is called a polarization.

Alternatively, we can choose complex coordinates with respect  
to $J_-$; then we have
$\Omega = \Omega_-^{(2,0)} +\bar{\Omega}_-^{(0,2)}$, and 
$\Omega_-^{(2,0)}$ is again a holomorphic symplectic
structure. Thus we can introduce the coordinates $\{Q^{a'},\bar Q^{\bar 
a'}, P^{a'},\bar P^{\bar a'}\}~a'=1\dots d_s$ such that
\beq
\Omega_-^{(2,0)} = d Q^{a'} \wedge d P^{a'}~,\qquad
\bar{\Omega}_-^{(0,2)} = d\bar Q^{\bar a'} \wedge d\bar P^{\bar a'}~.
\eeq{complckdoe22}
In these coordinates $J_-$ has the form
\beq
J_- = \left(\begin{array}{cccc}
i & 0 & 0 &  0\\
0 & -i & 0 & 0 \\
0 & 0&  i &  0 \\
0 & 0 & 0& -i
\end{array}\right) \equiv \left( \begin{array}{cc}
J_s & 0\\
0 & J_s
\end{array} \right)~.
\eeq{form23-a---}
The coordinate transformation $\{q,p\}\to\{Q,P\}$  
preserves $\Omega$, and hence is a canonical transformations
(symplectomorphisms). A canonical transformation can always be
described by a generating function $K$ that depends a $d_s$-dimensional
subset of the ``old'' coordinates $\{q,p\}$ and a $d_s$-dimensional
subset of the ``new'' coordinates $\{Q,P\}$ (see, \eg, \cite{arnold}).
For simplicity, we choose our polarization such that the  
generating function $K$ depends on the ``old'' $q$ and 
the ``new'' $P$ coordinates; it is a theorem that such a polarization always
exists \cite{arnold}. 

Thus in a neighborhood, the canonical transformation is
given by the generating function $K(q,P)$
\beq
p = \frac{\d K}{\d q}~,\qquad Q = \frac{\d K}{\d P}~.
\eeq{denegaiwp298}
We now calculate $J_+$, $J_-$ and $\Omega$ in the ``mixed''
coordinates $\{q, P\}$.
Consider $J_+$.  In $\{q, P\}$ coordinates $J_+$ is  
given by
\beq
J_+ = \left(\frac {\d (q, p)}{\d (q, P)} \right)^{-1} \left(\!\begin{array}{ll}
J_s & 0\\
0 & J_s
\end{array}\! \right) \! \left( \frac {\d (q, p)}{\d (q, P)}\right) .
\eeq{transformayiwe}
The transformation matrix is given as
\beq
\frac {\d (q, p)}{\d (q, P)}  = \left( \begin{array}{cc}
1 &  0 \\
\frac{\d p}{\d q} & \frac{\d p} {\d P}
\end{array}
\right)  = \left( \begin{array}{cc}
1 &  0 \\
\frac{\d^2 K}{\d q \d q} & \frac{\d^2 K} {\d P\d q}
\end{array}
\right) \equiv \left( \begin{array}{cc}
1 & 0 \\
K_{LL} & K_{LR}
\end{array}
\right)
\eeq{tarsnforamslwo}
where in complex coordinates we have
\beq
K_{LL} = \left( \begin{array}{cc}
K_{ab} & K_{a\bar{b}} \\
K_{\bar{a}b} & K_{\bar{a}\bar{b}}
\end{array}\right)~,\qquad
K_{LR} = \left( \begin{array}{ll}
K_{ab'} & K_{a\bar{b'}} \\
K_{\bar{a}b'} & K_{\bar{a}\bar{b'}}
\end{array}\right),
\eeq{definsoqwA}
and we have anticipated our identification the generating function $K(q,P)$ 
with the action $K(\mathbb{X}_L,\mathbb{X}_R)$ by introducing the labels $R,L$.
We find
\beq
J_+ = \left( \begin{array}{cc}
1 & 0 \\
-K^{-1}_{RL}K_{LL} & K_{RL}^{-1}
\end{array}
\right)\!
\left(\begin{array}{cc}
J_s & 0\\
0 & J_s
\end{array} \right)\!
\left( \begin{array}{cc}
1 & 0 \\
K_{LL} & K_{LR}
\end{array}
\right) = \left( \begin{array}{cc}
J_s & 0\\
K_{RL}^{-1} C_{LL} & K_{RL}^{-1}J_s K_{LR}
\end{array} \right),
\eeq{tasrau29393}
where $K_{LR}$ and $C_{LL}$ are defined in (\ref{commnotation}) in
terms of second derivatives of the generating function $K$. 
Thus in the coordinates $\{q,P\}$,  $J_+$ is given by (\ref{tasrau29393}).
Identifying the generating function $K(q,P)$ 
with the action $K(\mathbb{X}_L,\mathbb{X}_R)$, this result coincides
with the one we get from the semichiral sigma 
model \cite{Buscher:1987uw,Bogaerts:1999jc}
(c.f. \enr{Jplus} with no chiral or twisted chiral fields.).

Next we calculate $J_-$ in $\{q,P\}$ coordinates
\beq
J_- = \left(\frac {\d (Q, P)}{\d (q, P)} \right)^{-1} \left 
(\begin{array}{cc}
J_s & 0\\
0 & J_s
\end{array} \right)\!\left( \frac {\d (Q, P)}{\d (q, P)}\right) ,
\eeq{transformayiwenew}
where
\beq
\frac {\d (Q, P)}{\d (q, P)}  = \left( \begin{array}{cc}
\frac{\d Q}{\d q} & \frac{\d Q} {\d P} \\
0 & 1
\end{array}
\right)  = \left( \begin{array}{cc}
\frac{\d^2 K}{\d q \d P} & \frac{\d^2 K} {\d P\d P}\\
0 &  1
\end{array}
\right) \equiv \left( \begin{array}{cc}
K_{RL}& K_{RR}\\
\,\,0 & \,\,1
\end{array}
\right) .
\eeq{tarsnforamslwonew}
In complex coordinates  $K_{RL}=(K_{LR})^t$ defined as in (\ref 
{definsoqwA})
and $K_{RR}$ is
\beq
K_{RR}= \left( \begin{array}{cc}
K_{a'b'} & K_{a'\bar{b'}}\\
K_{\bar{a'}b'} & K_{\bar{a'}\bar{b'}}
\end{array} \right).
\eeq{definD1738}
Thus we can rewrite (\ref{transformayiwenew}) as
\beq
J_- = \left(\begin{array}{cc}
K_{LR}^{-1} & -K_{LR}^{-1}K_{RR} \\
0 & 1
\end{array} \right)\! \left(\begin{array}{cc}
J_s & 0\\
0 & J_s
\end{array} \right)\! \left( \begin{array}{cc}
K_{RL} & K_{RR} \\
0 & 1 \end{array} \right) = \left(\begin{array}{cc}
K_{LR}^{-1}J_sK_{RL} &  K_{LR}^{-1}C_{RR} \\
0 & J_s \end{array} \right),
\eeq{defksow292}
where $C_{RR}$ was defined in (\ref{commnotation}).
Once more, we have reproduced the semichiral expression (c.f. \enr{Jminus}).

Finally $\Omega$ in coordinates $(q,P)$ is given by
\beq
\Omega = \left(\begin{array}{cc}
0 & K_{LR} \\
- K_{RL} & 0
\end{array} \right).
\eeq{defkspoap}
In these coordinates the metric $g$ is given by \cite{Bogaerts:1999jc}
\beq
g = \Omega [J_+, J_-]
\eeq{definsmetric}
and this is the same as from semichiral considerations. 

Thus we have shown that the metric can be expressed in terms of second derivatives of
a single potential $K$. However, unlike the case of standard K\"ahler geometry,
the metric is not linear in the derivatives of $K$.  It is natural
to refer to $K$ as a generalized K\"ahler potential. This potential
has the interpretation simultaneously as a superspace Lagrangian and
as the generating function of a canonical transformation\footnote{This 
situation was found previously for $N=(4,4)$ hyperk\"ahler 
sigma models described in projective superspace \cite{LR}.}
between the complex coordinates adapted to $J_+$ and the 
complex coordinates adapted to $J_-$.

Furthermore, recalling that we have assumed 
${\rm ker}[J_+,J_-]=\emptyset$ throughout this section, the form $(\Omega^{(2,0)})^{d_s}$ is 
nondegenerate and defines a holomorphic volume form. 
Thus this is a generalized Calabi-Yau manifold \cite{hitchinCY}.

Finally, one may wonder if there actually exist examples where 
$\ker [J_ +, J_-] = \emptyset$. The work of \cite{Buscher:1987uw} provides
a local example in four-dimensions; in arbitrary dimensions, one can
consider hyperk\"ahler manifolds: 

\begin{theorem}
A generalized K\"ahler manifold with the
anticommutator of $J_+$ and $J_-$ constant, \ie, $\{ J_ +, J_-\}=c \mathbb{I}$,
is a hyperk\"ahler manifold whenever $|c|<2$.
\end{theorem}
{\it Proof:~} Using (\ref{nablJH}), the proof is straightforward in  
local coordinates. Alternatively one can observe that $B= \Omega \{ J_+, J_-\}$
\cite{Bogaerts:1999jc}, and hence the torsion, which is proportional to $dB$, vanishes. 
The explicit complex structures of the hyperk\"ahler manifold can be chosen as:
\be
I = J_+~,~~~ 
J =\frac1{\sqrt{1-\frac{c^2}4}}\left(J_- + \frac{c}2\,J_+\right)~,~~~ 
K=IJ~.
\ee
The construction we have presented can be applied to the hyperk\"ahler  
case with a new generalized
K\"ahler potential. Indeed from the condition $\{ J_+, J_-\}=c \mathbb{I}$,  
we get a partial differential equation for $K$ in the hyperk\"ahler  
case. In \cite{Sevrin:1996jr} it has been pointed
that in four dimensions, for $c=0$, this is the Monge-Amp\`ere equation.

\section{General case}
\label{GENERAL}

We now turn to the general case with both ker($[J_+,J_-]$) 
and coker($[J_+,J_-]$) notrivial. Essentially, we have to combine the  
arguments presented in the two previous sections.

We assume that in a neighborhood of $x_0$, the ranks of $\pi_\pm$ are  
constant, and as result, the rank of $\sigma$ is constant.  We   
work in coordinates adapted to the symplectic foliation of $\sigma$.  
Combining the notations from previous sections, we can chose coordinates $\{q,p,z,z'\}$ in  
which $J_+$ has the canonical form
\beq
J_+ = \left(\begin{array}{cccc}
J_s & 0 & 0& 0 \\
0 & J_s  & 0 & 0\\
0 & 0 & J_c & 0\\
0 & 0 & 0 &  J_t
\end{array} \right),
\eeq{canoniicalfromforJ}
where we use the notation (\ref{defincanoncsysm}).
The coordinates $z$ and $z'$ parametrize the kernels of $\pi_\mp$, respectively.
Thus $\{z,z'\}$ parametrize  the kernel of $\sigma$ and $\{q,p\}$  
are the Darboux coordinates for a symplectic leaf. On a leaf the symplectic  
form is given by (\ref{drholomcoord}). Alternatively we can choose the  
coordinates $\{Q, P, z, z'\}$ in which $J_-$ has a canonical 
form\footnote{We chose signs that are consistent 
with the sigma model results.}
\beq
J_- = \left(\begin{array}{cccc}
J_s & 0 & 0& 0 \\
0 & J_s  & 0 & 0\\
0 & 0 & J_c & 0\\
0 & 0 & 0 & - J_t
\end{array} \right).
\eeq{canoniicalfromforJextra}
Again $(Q, P)$ are the Darboux coordinates on a leaf with the  
symplectic form given by
(\ref{complckdoe22}).  If we fix a leaf (\ie, put $(z, z')$ to  
a fixed value) then we can apply
the discussion from Section \ref{COKER}.  Thus we can choose   
new coordinates $\{q, P\})$ along a leaf in a neighborhood of
$(q_0, p_0)$   (see the discussion of the existence
of these coordinates in Section \ref{COKER}). There exists a  
generating function $K$
such that the relations (\ref{denegaiwp298}) are satisfied.   
This argument can be a applied
to a single leaf. If we change to another leaf then we get  
another generating function.
Thus in a neighborhood of $x_0$ we have a family\footnote 
{One may wonder if the dependence
of $K$ on $z$ and $z'$ is smooth; this is necessary 
to write the coordinate transformation to $\{q, P, z, z'\}$. 
The existence of these coordinates follows from Arnold's result
 \cite{arnold}.}
of generating functions $K(q, P, z, z')$
such that
\beq
p =\frac{\d K}{\d q}~,\qquad Q=\frac 
{\d K}{\d P}
\eeq{impiroao3994}
is satisfied. With this definition,
$K(q, P, z, z')$ is defined up to the addition of an arbitrary function $f(z, z')$.

Now we can calculate $J_\pm$ in the coordinates $\{q, P, z, z'\}$; the complex 
structure $J_+$ is 
\beq
J_+ = \left( \frac{\d (q, p, z, z')}{\d(q, P, z, z')} \right) 
^{-1} \left(\begin{array}{cccc}
J_s & 0 & 0& 0 \\
0 & J_s  & 0 & 0\\
0 & 0 & J_c & 0\\
0 & 0 & 0 &  J_t
\end{array} \right)
\left( \frac{\d (q, p, z, z')}{\d(q, P, z, z')} \right).
\eeq{defdlfp384940}
The transformation matrix is given as
\bea
\frac{\d (q, p, z, z')}{\d(q, P, z, z')}  ~=~ \left( \begin 
{array}{cccc}
1 & 0 & 0 & 0 \\
\frac{\d p}{\d q} & \frac{\d p}{\d P} & \frac{\d p}{\d z} &  
\frac{\d p}{\d z'}\\
0 & 0 & 1&0\\
0 & 0 & 0 & 1
\end {array} \right)
&=& \left( \begin{array}{llll}
1 & 0 & 0 & 0 \\
\frac{\d^2 K}{\d q \d q} & \frac{\d^2 K}{\d P\d q} & \frac 
{\d^2 K}{\d z\d q} & \frac{\d^2 K}{\d z' \d q}\\
0 & 0 & 1&0\\
0 & 0 & 0 & 1
\end {array}\! \right)\qquad\cr\cr\cr
&= &\,\left( \begin{array}{cccc}
1 & 0 & 0 & 0 \\
K_{LL} & K_{LR} & K_{Lc} & K_{Lt}\\
0 & 0 & 1&0\\
0 & 0 & 0 & 1
\end {array} \right)~,
\label{tardheosp3e90}
\eea
where in complex coordinates $K_{LL}$ and $K_{LR}$ were defined in  
(\ref{definsoqwA}) and
\beq
K_{Lc} = \left(\begin{array}{cc}
K_{a\alpha} & K_{a\bar{\alpha}}\\
K_{\bar{a}\alpha} & K_{\bar{a}\bar{\alpha}}
\end{array} \right),\qquad K_{Lt} = \left(\begin{array}{cc}
K_{a\alpha'} & K_{a\bar{\alpha}'}\\
K_{\bar{a}\alpha'} & K_{\bar{a}\bar{\alpha}'}
\end{array} \right).
\eeq{defirnowo038983903}
Next using (\ref{defdlfp384940}) and (\ref{tardheosp3e90}) we  
calculate $J_+$
\beq
J_+ = \left( \begin{array}{cccc}
J_s & 0 & 0 &  0\\
K^{-1}_{RL}C_{LL} & K_{RL}^{-1}J_s K_{LR} &  K_{RL}^{-1}C_{Lc} &
K_{RL}^{-1}C_{Lt} \\
0 & 0 & J_c & 0 \\
0 &0 & 0 & J_t
\end{array} \right),
\eeq{fullJallcasesl}
where all of the $C$ matrices are defined in (\ref{commnotation}).
This is exactly the same expression one gets from the sigma model
considerations (\ref{Jplus}).

Similarly, we calculate the form of $J_-$ in $\{q, P, z, z'\}$  
coordinates:
\beq
J_- = \left( \frac{\d (Q, P, z, z')}{\d(q, P, z, z')} \right) 
^{-1} \left(\begin{array}{cccc}
J_s & 0 & 0& 0 \\
0 &  J_s  & 0 & 0\\
0 & 0 & J_c & 0\\
0 & 0 & 0 &  -J_t
\end{array} \right)
\left( \frac{\d (Q, P, z, z')}{\d(q, P, z, z')} \right).
\eeq{defdlfp384940extra}
\beq
J_-= \left( \begin{array}{cccc}
K_{LR}^{-1}J_s K_{RL}  & K_{LR}^{-1}C_{RR} & - K_{LR}^{-1}C_{Rc} &
K_{LR}^{-1}A_{Rt} \\
0 &  -J_s & 0 & 0\\
0 & 0 & J_c & 0 \\
0 &0 & 0 & -J_t
\end{array} \right),
\eeq{fullJallcaseslextra}
where again the $C$ and $A$ matrices were defined in (\ref 
{commnotation})
and $K_{Rc}$ and $K_{Rt}$ are
\beq
K_{Rc} = \left(\begin{array}{cc}
K_{a'\alpha} & K_{a'\bar{\alpha}}\\
K_{\bar{a}'\alpha} & K_{\bar{a}'\bar{\alpha}}
\end{array} \right),\qquad
K_{Rt} = \left(\begin{array}{cc}
K_{a'\alpha'} & K_{a'\bar{\alpha}'}\\
K_{\bar{a}'\alpha'} & K_{\bar{a}'\bar{\alpha}'}
\end{array} \right).
\eeq{defirnowo038983903extra}
This is exactly the same expression one gets from the sigma model \enr{Jminus}.

We now consider the metric;  in the coordinates $\{q, P,  z, z'\}$,
 the metric has a form
\beq
g = \left( \begin{array}{llll}
g_{AB} & g_{AB'} & g_{A{\cal B}} & g_{A\cal{B}'} \\
g_{A'B} & g_{A'B'} & g_{A'{\cal B}} & g_{A'\cal{B}'} \\
g_{{\cal A}B} & g_{{\cal A}B'} & g_{{\cal A}{\cal B}} & g_ 
{{\cal A}\cal{B}'} \\
g_{{\cal A}'B} & g_{{\cal A}'B'} & g_{{\cal A}'{\cal B}} & g_ 
{{\cal A}'\cal{B}'}
\end{array} \right).
\eeq{formameudoa}
The definition (\ref{definHitchpoisos}) of the Poisson structure
$\sigma$ determines all the components of the metric $g$ {\it except}
those along the kernel of $\sigma$: $g_{{\cal A}{\cal B}},
g_{{\cal A}\cal{B}'},  g_{{\cal A}'{\cal B}}, g_{{\cal A}'\cal 
{B}'}$; this matches the ambiguity in the generating function $K(q,P,z,z')$
noted below \enr{impiroao3994}. The remaining components of
the metric can be expressed in terms of the second  
derivatives of $K(q, P, z, z')$ using the relation (\ref{veryusefulrelapdkjas}):
\beq
J^\lambda_{+\mu} J^\sigma_{+\nu} J^\gamma_{+\rho}
(d\omega_+ )_{\lambda\sigma\gamma} =
- J^\lambda_{-\mu} J^\sigma_{-\nu} J^\gamma_{-\rho}
(d\omega_- )_{\lambda\sigma\gamma}~.
\eeq{fulldfefldkal3}
This is obvious in the K\"ahler case
($J_+ = J_-$), and was shown to be true whenever the
$[J_ +, J_-]=0$ in \cite{Gates:1984nk}.
In the general case, we argue as follows:
choosing  the local  coordinates $(q, P, z,  
z')$ we can plug the complex structures
(\ref{fullJallcasesl}) and (\ref{fullJallcaseslextra}) into (\ref 
{fulldfefldkal3}). After this the relation
(\ref{fulldfefldkal3}) becomes a first order partial  
differential equation for the metric $g$.
The differential equation contains the derivatives of $K$.  
However, we know a solution
for $g$ (which is indeed expressible completely in terms of the second  
derivatives of $K$):  it is precisely the expression derived
from the sigma model (see the expression for $E$ in Section  
\ref{OFF}). Similarly, (\ref{veryusefulrelapdkjas}) can be used to determine
the 2-form $B$ in terms of the second derivatives of $K$.

Thus we have established the existence of a generalization of  
the concept of a K\"ahler potential for generalized K\"ahler geometry.
It is natural to refer to this function as a generalized 
K\"ahler potential. Of course, as we found in the previous section,
the second derivatives of the generalized K\"ahler potential appear  
nonlinearily in the metric.
\section{Summary and discussion}
\label{SUMMARY}

We have resolved the long standing problem of finding manifestly 
off-shell supersymmetric formulation
for the general $N=(2,2)$ sigma model.
We have shown that the full set of fields which
is necessary for the description of general $N=(2,2)$ sigma model
consists of chiral, twisted chiral, and semichiral fields.
At the geometrical level this implies important
results about the generalized K\"ahler geometry, in particular the  
existence of a generalized K\"ahler potential. 
Thus for generalized K\"ahler  manifold
all the differential geometry can be locally encoded in a single  
real function. We have presented a
geometrical proof of this which is essentially independent of  
sigma model considerations. The only assumption we made was the   
regularity of the Poisson structures $\pi_\pm$ in a given neighborhood;
presumably, continuity allows one to relax this assumption in most
cases of physical interest. In general, it would be interesting
to go beyond this assumption; this would require the  
full apparatus of Poisson geometry, in particular a study of the  
transversal Poisson structures around $x_0$.

It follows that one can now discuss the general $N=(2,2)$ sigma  
models entirely within the powerful $N=(2,2)$
superfield formalism. In particular such problem as finding  
quotients of generalized K\"ahler
manifolds can be studied in all generality in this formalism.  
We plan to come back to this elsewhere.

From the mathematical point of view, it would be interesting to  
systematically study the first order PDE for the metric that arises from the
equation (\ref{fulldfefldkal3}).  Taking into account the  
discussion in Section \ref{COKER}, we seem to have some new
tools with which to study hyperk\"ahler manifolds.

\bigskip\bigskip
\noindent{\bf\Large Acknowledgement}:
\bigskip

\noindent We are grateful to the 2005 Simons Workshop
for providing the stimulating atmosphere where this work was initiated.
UL supported by EU grant (Superstring theory)
MRTN-2004-512194 and VR grant 621-2003-3454.
The work of MR was supported in part by NSF grant no.~PHY-0354776
and Supplement for International Cooperation with Central and Eastern  
Euorpe PHY 0300634. The research of R.v.U. was supported by 
Czech ministry of education contract No. MSM0021622409 and 
by Kontakt grant ME649. The research of M.Z. was
supported by VR-grant 621-2004-3177 and in part by the National Science  
Foundation under the Grant No. PHY99-07949.

\appendix
\Section{$N=(1,1)$ supersymmetry}
\label{a:11susy}

In this and the next appendix we collect our notation for $N=(1,1)$  
and $N=(2,2)$ superspace. In our conventions we closely 
follow \cite{Hitchin:1986ea}.

We use real (Majorana) two-component spinors $\psi^\alpha=
(\psi^+, \psi^-)$. Spinor indices are raised and lowered with the  
second-rank antisymmetric symbol $C_{\alpha\beta}$, 
which defines the spinor inner product:
\beq
C_{\alpha\beta}=-C_{\beta\alpha}=-C^{\alpha\beta}~,\qquad C_{+-} 
=i~,\qquad
\psi_\alpha =\psi^\beta C_{\beta\alpha}~,\qquad \psi^\alpha= C^ 
{\alpha\beta} \psi_\beta~.
\eeq{Cdef}
Throughout the paper we use  $(\+,=)$ as worldsheet indices, and $ 
(+,-)$ as two-dimensional spinor
indices.  We also use superspace conventions where the pair of spinor
coordinates of the two-dimensional superspace are labelled $\th^{\pm}$,
and the spinor derivatives $D_\pm$ and supersymmetry generators
$Q_\pm$ satisfy
\ber
D^2_+ &=&i\d_\+~, \qquad
D^2_- =i\d_=~, \qquad \{D_+,D_-\}=0~,\cr
Q_\pm &=& iD_\pm+ 2\th^{\pm}\d_{\pp}~,
\eer{alg}
where $\d_{\pp}=\partial_0\pm\partial_1$. 
The supersymmetry transformation of a superfield $\P$ is given by
\ber
\delta \P &\equiv &-i(\e^+Q_++\e^-Q_-)\P \cr
&=& (\e^+D_++\e^-D_-)\P
-2i(\e^+\th^+\d_\++\e^-\th^-\d_=)\P ~.
\eer{tfs}
The components of a scalar superfield $\P$ are defined by  
projection as follows:
\ber
\P|\equiv X~, \qquad D_\pm\P| \equiv \p_\pm~, \qquad D_+D_-\P|\equiv F~,
\eer{comp}
where the vertical bar $|$ denotes ``the $\th =0$ part''.
The $N=(1,1)$  spinorial measure is conveniently written in terms of spinor derivatives:
\beq
\left.\int d^2\theta \,\,{\cal L} =   (D_+ D_- {\cal L})\right| .
\eeq{sssssssssspp}

\Section{$N=(2,2)$ supersymmetry}
\label{a:22susy}

In $N=(2,2)$ superspace, we have two independent $N=(1,1)$ subalgebras with spinor
derivatives $D_\alpha^1,D_\alpha^2$; 
we define complex complex spinor derivatives
\beq
\mathbb{D}_\alpha\equiv \frac{1}{2}(D_\alpha^1+iD_\alpha^2)~,\qquad
\bar{\mathbb{D}}_\alpha =\frac{1}{2}(D_\alpha^1 - iD^2_\alpha)
\eeq{compspderdef}
which obey the algebra
\beq
\begin{array}{ll}
\{\mathbb{D}_+, \bar{\mathbb{D}}_+ \} = i\d_\+~,~~ &
\{ \mathbb{D}_-, \bar{\mathbb{D}}_- \} = i\d_= ~,~~\\
\{\mathbb{D}_\alpha, \mathbb{D}_\beta\}= 0~, ~~ &\{ \bar{\mathbb{D}}_ 
\alpha, \bar{\mathbb{D}}_\beta \} =0 ~.~~
\end{array}
\eeq{compder}
These can be written in terms of complex spinor coordinates:
\beq
\mathbb{D}_{\pm} = \d_{\pm}+ \frac{i}{2} \bar{\theta}^{\pm} \d_ 
{\pp}~,\qquad
\bar{\mathbb{D}}_{\pm} = \bar{\d}_{\pm}+ \frac{i}{2} \theta^{\pm}  
\d_{\pp}~.
\eeq{DDbarDD}
In terms of the covariant derivatives, the supersymmetry  
generators  are
\beq
\mathbb{Q}_{\alpha}= i\mathbb{D}_{\alpha} + \theta^\beta \d_{\alpha 
\beta}~,\qquad
\bar{\mathbb{Q}}_{\alpha} = i\bar{\mathbb{D}}_\alpha +\bar{\theta}^ 
\beta \d_{\alpha\beta}~.
\eeq{QbarQ}
The supersymmetry transformation of a superfield
$\Phi$ is then defined by
\beq
\delta \Phi = i(\epsilon^\alpha \mathbb{Q}_\alpha + \bar{\epsilon}^ 
\alpha \bar{\mathbb{Q}}_\alpha)\Phi~.
\eeq{susyvarD}
Irreducible representations of $N=(2,2)$ obey constraints that are compatible
with the algebra \enr{compder}; for example,  
a chiral superfield ($\bar \mathbb{D}_{\pm} \Phi =0$)  
has components defined via projections as follows
\ber
\P|\equiv X~, \qquad \mathbb{D}_\pm\P| \equiv \p_\pm~, \qquad \mathbb{D}_+ 
\mathbb{D}_-\P|\equiv F~,
\eer{compcomp}
and a twisted chiral superfield ($\bar\mathbb{D}_+\chi=\mathbb{D}_-\chi=0$)
has components:
\ber
\chi|\equiv \tilde X~, \qquad \mathbb{D}_\+\chi| \equiv\tilde \psi_+~,
 \qquad \bar\mathbb{D}_-\chi| \equiv\tilde \psi_-~, \qquad \mathbb{D}_+ 
\bar\mathbb{D}_-\chi|\equiv \tilde F~,
\eer{compchi}
The $N=(2,2)$  spinorial measure is conveniently written in terms of spinor derivatives:
\beq
\left.\int d^2\theta\,d^2\bar{\theta} \,\,{\cal L} =  (\mathbb{D}_+ \mathbb 
{D}_- \bar{\mathbb{D}}_+
\bar{\mathbb{D}}_-  {\cal L})\right|.
\eeq{spinmesu}

\Section{Poisson geometry}
\label{a:poisson}

A ($d$-dimensional) manifold $M$ is Poisson if it admits an antisymmetric bivector
$\pi \in \wedge^2 TM$
that satisfies the differential condition
\beq
\pi^{\mu\nu} \d_\nu \pi^{\rho\sigma} +\pi^{\rho\nu} \d_\nu \pi^ 
{\sigma\mu}
+\pi^{\sigma\nu} \d_\nu \pi^{\mu\rho}  =0~.
\eeq{differancondPP}
If $\pi$ is invertible, $\pi^{-1}$ is a symplectic structure.
The bivector $\pi$ defines the conventional Poisson bracket
\beq
\{ f, g \} \equiv \pi(df, dg) = \pi^{\mu\nu} \d_\mu f \, \d_\nu g~,\qquad f(x), g(x)
\in C^\infty(M)~,
\eeq{poissbr}
which is a bilinear map $C^\infty(M)\times C^\infty(M)
\rightarrow C^\infty(M)$. Because of (\ref{differancondPP}), the
Poisson bracket (\ref{poissbr}) has the ordinary antisymmetry property
and satisfies the standard Leibnitz rule and Jacobi identity.

Next we recall that (locally) a Poisson manifold admits a foliation by
symplectic leaves.
Let $M$ be a Poisson manifold
with the Poisson structure $\pi^{\mu\nu}$, $\mu,\nu =1,2,...,d$; 
choose a point $x_0$ such that in its neighborhood $\rank(\pi)=n$
is constant. Such a point is called regular.\footnote{In general, 
a non-regular Poisson manifold has
singular points where the rank jumps \cite{vaisman}.
We do not discuss these points and their neighborhoods here.}

A vector field is locally Hamiltonian if it can be written
as the contraction of the bivector $\pi$ with a closed
one-form $e$ (locally $e=df$ for some function $f$).
The Lie bracket of two locally Hamiltonian vector fields 
is again locally Hamiltonian:
\beq 
{\rm for}~v^\mu_A\equiv \pi^{\mu\nu} \d_\nu f_A~,~~~
({\cal L}_{v_A}v_B)^\mu\equiv v^\nu_A\d_\nu v^\mu_B-v^\nu_B\d_\nu v^\mu_A=
\pi^{\mu\rho} \d_\rho \left( (\d_\nu f_B) \pi^{\nu 
\lambda}(\d_\lambda f_A)\right) . 
\eeq{setHMF}
The maximum number of linearly independent 
locally Hamiltonian vector fields in the neighborhood of 
a regular point $x_0$ is clearly $n=\rank(\pi)$; 
then Frobenius theorem implies
that the vector fields locally generate an integral submanifold $S$ through 
$x_0$, and it is always possible to introduce the local coordinates  
$x^\mu=\{x^A, x^i\}$, $A=1,\dots,n$, $i=n+1, ..., d$ in the
neighborhood of $x_0$ such that $S$ can be described by 
$x^i=constant$ and $x^A$ are the coordinates on $S$.  
The restriction of the Poisson bracket to
the functions on the submanifold $S$ is again a Poisson bracket, and  
is indeed a {\em nondegenerate} Poisson structure on $S$. 
As a result, in the coordinates $x^\mu= \{x^A, x^i\}$, 
$\pi$ has the following form
\beq 
\pi^{\mu\nu} = \left( \begin{array}{cc} \pi^{AB} & 0 \\
0 & 0 \end{array} \right).
\eeq{formP}
Since $\pi^{AB}\equiv\pi|_S$ is nondegenerate, 
it is the inverse of a symplectic structure on $S$, and
thus the Poisson manifold is foliated by
symplectic leaves.  In a generic coordinate system, there is a  
locally complete set of $d-n$ independent Casimir functions $\{f_i(x)\}$ of $\pi$ which have
vanishing Poisson bracket with any function from $C^\infty({\cal M})$. In
these coordinates the symplectic leaves are determined locally by the
conditions $f_i(x)=constant.$

For further details on the Poisson geometry the reader may consult  
the  book \cite{vaisman}.

\eject


\begin{thebibliography}{6666}

\newcommand{\np}{{\em Nucl.\ Phys.\ }}
\newcommand{\pr}{{\em Phys.\ Rev.\ }}
\newcommand{\cmp}{{\em Commun.\ Math.\ Phys.\ }}
\newcommand{\pl}{{\em Phys.\ Lett.\ }}
%
\bibitem{Gates:1984nk}
S.~J.~.~Gates, C.~M.~Hull and M.~Ro\v{c}ek,
``Twisted Multiplets And New Supersymmetric Nonlinear Sigma Models,''
Nucl.\ Phys.\ {\bf B248} (1984) 157.
%%CITATION = NUPHA,B248,157;%%
\bibitem{Buscher:1987uw}
T.~Buscher, U.~Lindstr\"om and M.~Ro\v{c}ek,
``New Supersymmetric Sigma Models With Wess-Zumino Terms,''
Phys.\ Lett.\ {\bf B202}, 94 (1988).
%%CITATION = PHLTA,B202,94;%%
\bibitem{Sevrin:1996jr}
A.~Sevrin and J.~Troost,
``Off-shell formulation of N = 2 non-linear sigma-models,''
Nucl.\ Phys.\ {\bf B492} (1997) 623
[arXiv:hep-th/9610102].
%%CITATION = HEP-TH 9610102;%%
\bibitem{Bogaerts:1999jc}
J.~Bogaerts, A.~Sevrin, S.~van der Loo and S.~Van Gils,
``Properties of semichiral superfields,''
Nucl.\ Phys.\ {\bf B562} (1999) 277
[arXiv:hep-th/9905141].
%%CITATION = HEP-TH 9905141;%%
\bibitem{Curtright:1984dz}
T.~L.~Curtright and C.~K.~Zachos,
``Geometry, Topology And Supersymmetry In Nonlinear Models,''
Phys.\ Rev.\ Lett.\  {\bf 53}, 1799 (1984).
%%CITATION = PRLTA,53,1799;%%
\bibitem{Howe:1985pm}
P.~S.~Howe and G.~Sierra,
``Two-Dimensional Supersymmetric Nonlinear Sigma Models With Torsion,''
 Phys.\ Lett.\ {\bf B148}, 451 (1984).
%%CITATION = PHLTA,B148,451;%%   
\bibitem{Lyakhovich:2002kc}
S.~Lyakhovich and M.~Zabzine,
``Poisson geometry of sigma models with extended supersymmetry,''
Phys.\ Lett.\ {\bf B548} (2002) 243
[arXiv:hep-th/0210043].
%%CITATION = HEP-TH 0210043;%%
\bibitem{hitchinCY}
N.~Hitchin,
``Generalized Calabi-Yau manifolds,'' Q. J. Math. {\bf 54} (2003),  
no. 3, 281
308, [arXiv:math.DG/0209099].
%
\bibitem{gualtieri}
M.~Gualtieri, ``Generalized complex geometry,'' Oxford University  
DPhil thesis,
[arXiv:math.DG/0401221].
%
\bibitem{Lindstrom:2004eh}
U.~Lindstr\"om,
``Generalized N = (2,2) supersymmetric non-linear sigma models,''
Phys.\ Lett.\ {\bf B587}, 216 (2004)
[arXiv:hep-th/0401100].
%%CITATION = HEP-TH 0401100;%%
\bibitem{Lindstrom:2004iw}
U.~Lindstr\"om, R.~Minasian, A.~Tomasiello and M.~Zabzine,
``Generalized complex manifolds and supersymmetry,''
Commun.\ Math.\ Phys.\  {\bf 257}, 235 (2005)
[arXiv:hep-th/0405085].
%%CITATION = HEP-TH 0405085;%%
\bibitem{Lindstrom:2004hi}
U.~Lindstr\"om, M.~Ro\v{c}ek, R.~von Unge and M.~Zabzine,
``Generalized Kaehler geometry and manifest N = (2,2) supersymmetric
nonlinear sigma-models,''
JHEP~{\bf 0507} (2005) 067
[arXiv:hep-th/0411186].
%%CITATION = HEP-TH 0411186;%%
\bibitem{Ivanov:1994ec}
I.~T.~Ivanov, B.~B.~Kim and M.~Ro\v{c}ek,
``Complex structures, duality and WZW models in extended  
superspace,''
Phys.\ Lett.\ {\bf B343} (1995) 133
[arXiv:hep-th/9406063].
%%CITATION = HEP-TH 9406063;%%
\bibitem{Sevrin:1996jq}
A.~Sevrin and J.~Troost,
``The geometry of supersymmetric sigma-models,''
[arXiv:hep-th/9610103].
%%CITATION = HEP-TH 9610103;%%
\bibitem{Grisaru:1997pg}
M.~T.~Grisaru, M.~Massar, A.~Sevrin and J.~Troost,
``The quantum geometry of N = (2,2) non-linear sigma-models,''
Phys.\ Lett.\ {\bf B412}, 53 (1997)
[arXiv:hep-th/9706218].
%%CITATION = HEP-TH 9706218;%%
\bibitem{hitchinP}
N.~Hitchin,
``Instantons, Poisson structures and generalized K\"ahler geometry,"
[arXiv:math.DG/0503432].
%
\bibitem{arnold}
V.~I.~Arnold,
``Mathematical methods of classical mechanics,"
Translated from the Russian by K. Vogtmann and A. Weinstein.
Second edition. Graduate Texts in Mathematics, 60.
Springer-Verlag, New York, 1989. xvi+508 pp.
%
\bibitem{LR}
U.~Lindstr\"om and M.~Ro\v cek, private communication, in preparation.
%
\bibitem{Hitchin:1986ea}
N.~J.~Hitchin, A.~Karlhede, U.~Lindstr\"om and M.~Ro\v{c}ek,
``Hyperk\"ahler Metrics And Supersymmetry,''
Commun.\ Math.\ Phys.\  {\bf 108} (1987) 535.
%%CITATION = CMPHA,108,535;%%
\bibitem{vaisman}
I.~Vaisman,
``Lectures on the Geometry of Poisson Manifolds'' Progress in  
Mathematics, Vol {\bf 118} (Birkh\"auser, Basel, 1994).
%
\end{thebibliography}
\end{document}